\documentstyle[11pt,paspconf]{article}

\input psfig
\newcommand{\etal}{{\sl et al.}}
\newcommand{\iras}{{\sl IRAS}}
\def\kms{$\rm km\,s^{-1}$}
\def\bfr{{\bf r}}

\begin{document}

\title{Questions and Controversies in the Measurement and
Interpretation of Large-Scale Flows}
\author{Michael A. Strauss}
\affil{Princeton University Observatory, Princeton, NJ 08544}

\begin{abstract}
This introductory talk to the 1999 Victoria Conference on Large-Scale
Flows will present the ``big questions'' which
will be discussed in the conference:
\begin{itemize} 
\item Does the velocity field converge on the largest scales?
\item Why can't we agree on the value of $\beta$? 
\item How can we properly measure the small-scale velocity dispersion?
\item Just how complicated can biasing be?
\item How universal are the distance indicators we are using?
\item How do we design our next generation of surveys to answer the
above questions?
\end{itemize}
\end{abstract}

\noindent One of the great advantages of giving a review talk at the
beginning of a conference is that I get to ask all the questions; it
is up to the rest of the participants to come up with the answers.
That being said, I will interject my own opinions here and there,
occasionally colored by what I learned at the conference itself (a
benefit of hindsight I of course did not have at the time I gave the
talk itself!).   However, I will resist the temptation
to steal too much from Avishai Dekel's conference summary.

\section{ The Large-scale Velocity Field}

If the universe approaches homogeneity as one looks on ever-larger
scales, then we expect the rms peculiar velocities to approach zero as
as we average over larger and larger spheres.  With this basic
notion in mind, astronomers since the path-breaking work of Vera Rubin
and co-workers in the mid-1970's have been trying to measure the scale
on which the bulk flow approaches zero, or equivalently, on which the
velocity field measured in the Local Group frame simply reflects
the 600 \kms\ motion of the Local Group with respect to the CMB.  A
definitive measurement of this scale would both confirm a fundamental
prediction of the Cosmological Principle and the gravitational
stability paradigm, and test our interpretation of the
dipole moment in the CMB.  Moreover, the measurement of the amplitude of the
flow as a function of scale is a strong probe of the matter power
spectrum on large scales; it is unaffected by bias, and is less
sensitive to non-linear effects, and more sensitive to the
longest-wavelength modes, than is the measurement of galaxy
fluctuations on equivalent scales (cf., Strauss 1997). 

With that in mind, Table 1 qualitatively lists some of the important
bulk flow measurements in the literature up to the time of this
meeting; apologies to any surveys I may have left out.  See Strauss \&
Willick (1995) for references to the older literature. 
\begin{table}
\caption{Recent Measurements of Large-Scale Bulk Flows}\label{table:1}
\begin{center}
\begin{tabular}{ll}
\tableline
Reference&Result\\
\tableline
\smallskip
Aaronson \etal\ (1986)&Convergence to Hubble flow at 6000 \kms\\
Lynden-Bell \etal\ (1988)&Flow towards the Great Attractor\\
Courteau \etal\ (1993)&Bulk flow of 300 \kms\ at 6000 \kms\\
Lauer \& Postman (1994)& Bulk flow of 600 \kms\ at 15,000 \kms\\
Riess \etal\ (1996)&SN are at rest with respect to the CMB\\
Giovanelli \etal\ (1998)&No bulk flow at 6000 \kms\\
Saglia \etal\ (1998; EFAR)&No bulk flow at 10,000 \kms\\
Hudson \etal\ (1999; SMAC)&Bulk flow of 600 \kms\ at 15,000 \kms,\\
&\quad in a different direction from LP\\
Willick (1999; LP10K)&Bulk flow of 600 \kms\ at 15,000 \kms,\\
&\quad in a different direction from LP\\
Dale \etal\ (1999)&No bulk flow to 20,000 \kms\\
\tableline
\end{tabular}
\end{center}
\end{table}

  Theoretical prejudice would suggest that on large scales, the bulk
flows should be of low amplitude, and a number of the analyses listed
in Table 1 indicate this.  The most direct challenge to this was that
of Lauer \& Postman (1994), who found a 6000 \kms\ flow within a
volume of 15,000 \kms, which was completely unexpected. Many of the
more recent surveys listed above, and discussed in this conference,
were carried out to test this startling result.  Unfortunately, none
have confirmed the LP bulk flow, but the SMAC and LP10K results of
Hudson \etal\ (1999) and Willick (1999), respectively, seem to find a
bulk flow of similar amplitude, on similar scale, in a {\em different}
direction.  It is worth emphasizing that despite many efforts, no-one
has found any serious errors in the LP data or any methodological
problem with their analysis.

The results in this table are presented graphically in Figure 1, based on
a similar figure in Postman (1995), which shows the dependence of
these results on scale. 
\begin{figure}
\centerline{\psfig{file=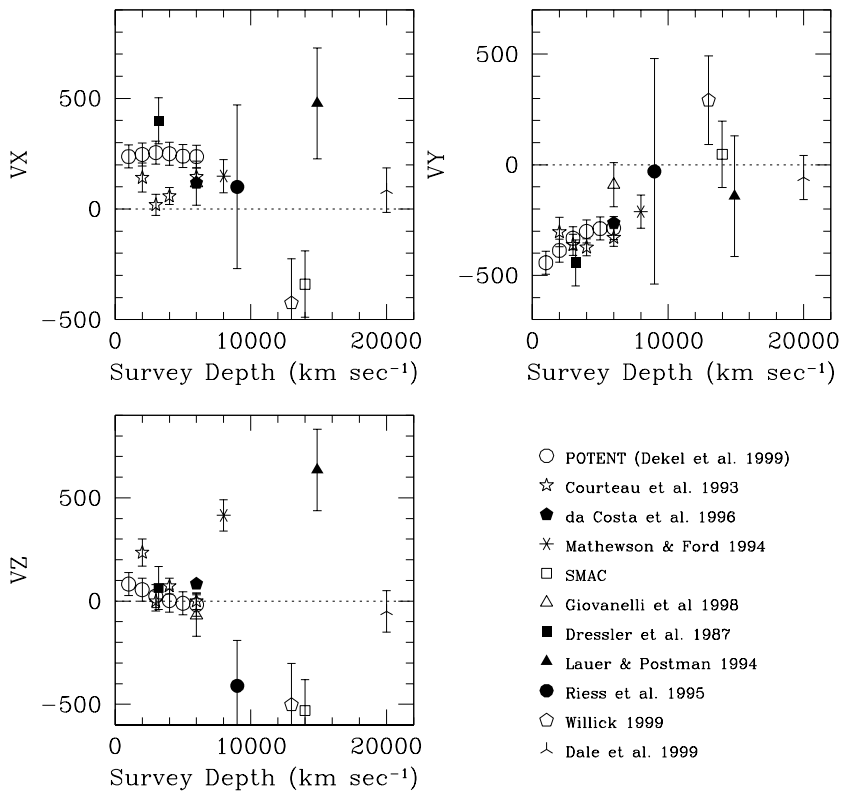,width=12cm}}
\caption{A compilation of bulk flow measurements as a function of
scale.   The
three panels give the components of the quoted bulk flows along the
Galactic $X$, $Y$, and $Z$ directions in \kms, as a function of the
depth of the various surveys. Error bars are as quoted by each paper,
and do not take into account the covariance between the different
directions (i.e., due to error ellipsoids whose principal axes are not
aligned with the Galactic Cartesian directions). Adapted from
Postman (1995).}
\label{fig:postman}
\end{figure}
Given the scatter in this diagram, we cannot yet claim as
a community to have come to an agreement as to the nature of the flows
on large scales.  But keep in mind that Figure~1 (and its variants,
which were presented throughout this meeting) is tremendously
misleading.  First, no single survey probes a given scale; the value
of the scale assigned to each data point is usually some ``typical'',
or weighted mean distance.  Second, the errors in the bulk flow are
really characterized by a $3\times 3$ covariance matrix, so the error
bars given here leave out a substantial amount of information.  Third,
and most important, because the velocity field is more complex than
just a bulk flow, any survey which is not perfectly uniformly sampled
across the sky with homogeneous errors will necessarily alias power
from smaller wavelengths in its bulk flow determination, a point made
most directly by Watkins \& Feldman (1995).  This means that one has
to take into account this aliasing in asking whether the bulk
flow results of two different surveys are in contradiction (see
Hudson's contribution to these proceedings).

Another warning, made by Marc Davis several times in this conference,
was that the error bars in such analyses often only reflect {\em
statistical} measurement errors (e.g., due to the known scatter of the
distance indicator used), and ignore {\em systematic} errors, due,
e.g., to the uncertainty in the calibration of the distance indicator
relation itself. These systematic errors need to be properly
quantified if we are to decide whether there are indeed serious
discrepancies between different measurements, as
Figure~\ref{fig:postman} appears to show. 

\section{The Value of $\mathbf \beta \equiv \Omega^{0.6}/b_{IRAS}$}

Linear perturbation theory gives a linear relation between the
peculiar velocity and gravity fields; alternatively, it can be
expressed as a linear relationship between the density field and the
{\em divergence} of the velocity field.  In the linear biasing
paradigm (see the discussion in \S~\ref{sec:bias}), the proportionality
factor is the quantity $\beta \equiv \Omega^{0.6}/b$. There is a very
extensive history of attempts to measure $\beta$ using a variety of
techniques comparing peculiar velocity and redshift survey data (see,
e.g., Strauss \& Willick 1995 for a review of the earlier literature).
I tabulate a rather incomplete list of such attempts in Table 2, with one
representative entry per method used, apologizing to those whose
results are left out.  To keep things simple, I've restricted myself
to recent analyses of \iras\ galaxies, to avoid questions of the
relative bias of \iras\ and optical galaxies (e.g., Baker \etal\
1998).  I have translated the results from the measured cluster
abundance to $\beta$ using the observed $\sigma_{8,g}$ of \iras\
galaxies (Fisher \etal\ 1994).

\begin{table}
\caption{Methods for Determining $\beta$}\label{table:2}
\begin{center}
\begin{tabular}{ll}
\tableline
Method/reference&Result\\
\tableline
\smallskip

$\delta_g$ versus $\nabla
\cdot {\bf v}$ &$\beta = 0.89 \pm 0.12$\\
\quad Sigad \etal\ (1998)\\ 
Maximum likelihood fit of velocity field model to TF data
&$\beta = 0.50 \pm 0.07$\\
\quad Willick \etal\ (1997,1998)\\ 
Expansion of velocity field in spherical harmonics &$\beta = 0.6 \pm
0.1$\\
\quad da Costa \etal\ (1998)\\  
Maximum likelihood fit of $P(k)$ to TF data &$\beta = 1.2 \pm
0.2$\\
\quad Freudling \etal\ (1999)\\ 
Anisotropy of redshift-space clustering &$\beta = 0.77 \pm
0.22$\\
\quad Hamilton (1998)\\ 
Anisotropy of spherical harmonic expansion of $\delta_g$ &$\beta =
0.58 \pm 0.26$\\
\quad Tadros \etal\ (1999)\\ 
Dipole moment of redshift surveys &$\beta = 0.6 \pm 0.15$\\
\quad Strauss \etal\ (1992)\\ 
Cluster abundance &$\beta = 0.7 \pm 0.1$\\
\quad Eke \etal\ (1996)\\ 
Depth of voids &$\Omega > 0.3$\\
\quad Dekel \& Rees (1994)\\
Gaussianity of velocity field &$\Omega = 1$\\
\quad Nusser \& Dekel (1993)\\
\tableline
\end{tabular}
\end{center}
\end{table}

\begin{figure}
\centerline{\psfig{file=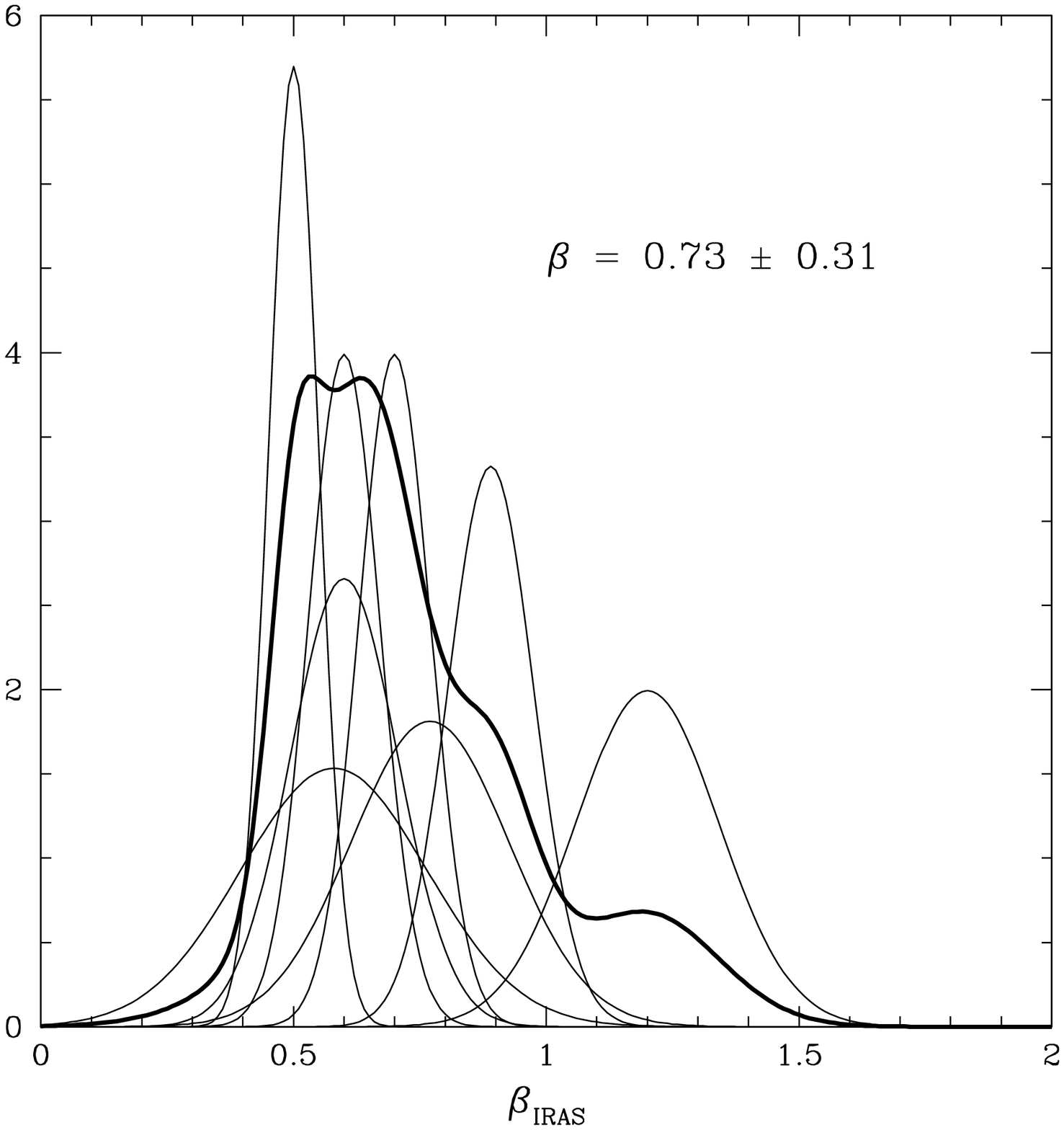,width=12cm}}
\caption{Each of the values in Table 2 is represented as a Gaussian
with mean and standard deviation as given.  The rescaled sum is the
heavy curve.\label{fig:beta}}
\end{figure}

Figure~\ref{fig:beta} plots each of
these determinations of $\beta$ as a Gaussian with mean and standard
deviation as given in the table.  
A qualitative sense of whether the community is
in agreement is whether the sum of these curves (shown, rescaled, as the
heavy curve) yields something sensible.  The answer is not very
convincing; the sum is far from a Gaussian, and is wide enough to
include values to make cosmologists of all persuasions happy.  It is
interesting to compare this with the equivalent plot in Strauss \&
Willick (1995), based on the literature up to that point; they find a
similarly broad sum, with the same mean (although not quite as bumpy),
although only a few of the datapoints are shared between the two
analyses. 

  Needless to say, this tells us that serious systematic effects are
influencing at least some of these analyses, especially considering
that many of them use the {\em same} underlying dataset, which means that the error
bars are very far from independent.  Among the likely important effects are:
\begin{itemize} 
\item The fact that the effective smoothing scales of these different
analyses are different; 
\item Nonlinear effects in the gravity;
\item A bias relation more complicated than linear deterministic bias
(see \S~5); 
\item Possible correlated and systematic errors in the data.
\end{itemize}

These issues are discussed in more detail in the panel discussion on
the value of $\beta$ (Willick \etal, these proceedings). 

\section{The Coldness of the Velocity Field}

Many workers over the past 15 years have remarked on the quietness of
the velocity field on small scales.   This result is tremendously
important for constraining
cosmological models.  It was the smallness of the small-scale velocity
dispersion that provided a first inkling to Davis \etal\ (1985) that
the standard $\Omega = 1$ CDM model was in trouble, and motivated them
to consider a model with $b = 2.5$ (which we now know to be in
dramatic contradiction to COBE normalization).  More generally, the
Cosmic Virial Theorem tells us that a small velocity dispersion
indicates a small value of $\Omega$. 

There are a number of ways in which the quietness of the velocity
field can be quantified:
\begin{itemize} 
\item Measuring the small-scale velocity dispersion by the anisotropy of the
galaxy correlation function in redshift space;
\item Dividing up the observed bulk flow into large-scale and
small-scale components;
\item Quantifying the residuals that remain after fitting the
large-scale flows to a smooth model. 
\end{itemize}

The small-scale velocity dispersion as measured from the
redshift-space correlation
function is a pair-weighted statistic, and is therefore heavily
weighted by the clusters in the survey volume.  However, the
small-scale velocity dispersion is a strong function of local density,
being quite a bit higher in the cores of clusters than in the field
(this is why redshift pie diagrams show dramatic fingers of God in
dense clusters, and thin filaments elsewhere), which means that this
statistic depends crucially on exactly how many clusters happen to
enter your survey volume (e.g., Mo \etal\ 1993).  A number of related
statistics have been suggested: working in $k$-space (Szalay \etal\
1998), measuring the small-scale velocity dispersion without the
pair-weighting (Davis \etal\ 1997; see also Baker, this conference),
and measuring the small-scale velocity dispersion as a function of
local density (Strauss \etal\ 1998).  The last two of these, at least,
indicate that the small-scale velocity dispersion in the field is {\em
very} small, less than 150 \kms, but not enough work has yet been done
to confirm whether this is consistent with, e.g., currently popular
cosmological models with $\Omega \approx 0.3$. 

Measuring the coldness of the velocity field from a redshift survey is
rather indirect; better is to work from a peculiar velocity survey directly.
Sandage (1986), Brown \& Peebles (1987), Burstein (1990) and others
have remarked on how quiet the directly observed velocity field is;
there seems to be little structure on scales smaller than the
large-scale bulk flows.  Groth \etal\ (1989) quantified this when they
subtracted a bulk flow from the observed peculiar velocity field, and
found essentially vanishing residuals.  More recently, the VELMOD
technique directly measures the amplitude of velocity field residuals,
after accounting for peculiar velocity measurement errors and the
\iras\ predicted velocity field model; Willick \etal\ (1997) and
Willick \& Strauss (1998) also find a small-scale noise below 150
\kms, and evidence that it is an increasing function of density. Tonry
\etal\ (1999 and this conference) find similar results from their
modeling of the flow field of the surface brightness fluctuation data. 

Again, it is not yet clear to me that the models are doing an adequate
job of matching this statistic, and and if not, what physical effects
might we be missing?  One important complicating effect is the
nature of the galaxy bias, which is not going to be simple on small
scales, and to which we now turn. 

\section{Galaxy Bias is Complicated, and Interesting}
\label{sec:bias}

Galaxy bias exists.  We know this from a number of observations:
the relative density of elliptical and spiral galaxies is a strong
function of environment (which tells us that it cannot be true that
{\em both} are unbiased relative to the dark matter), and the strong
clustering of Lyman-break galaxies at high redshift can only be
understood if they are strongly biased.  

Once upon a time, we parameterized our ignorance about the
relative distribution of dark matter and galaxies with the linear bias
model:
\begin{equation} 
\delta_g(\bfr) = b\,\delta(\bfr),
\label{eq:linear-bias} 
\end{equation}
where $b$ was implicitly assumed to be independent of the scale on
which $\delta$ was defined.  On small
scales, where the characteristic fluctuations in $\delta$ are
appreciably larger than unity, equation~(\ref{eq:linear-bias}) cannot strictly hold true for $b >
1$, as both $\delta$ and $\delta_g$ are bounded from below by $-1$.
More generally, we now believe that bias is quite a bit more
complicated a beast than equation~(\ref{eq:linear-bias}) would imply.  On
very large scales, there are compelling analytic arguments (e.g.,
Coles 1993; Scherrer \& Weinberg 1998) that $b$ should be independent
of scale.  However, on smaller
scales, things can be quite a bit more complicated.  In
particular, 
\begin{itemize} 
\item The galaxy density field can be a non-linear function of the
galaxy density field. 
\item There can be physical quantities other than the local density
which affect the formation of galaxies, meaning that there will
necessarily be some scatter (sometimes confusingly called
``stochasticity'') around any mean relation between the galaxy and
mass density field.
\item The bias function can depend on scale, especially on small
scales.  
\item The bias function is already known to depend on cosmological epoch.
The continuity equation generically predicts that bias should approach
unity with time (in the absence of galaxy formation and merging). 
\item The bias function is already known to depend on the sample of
galaxies; clustering strength depends on surface brightness,
luminosity (both optical and infrared), star-formation rate, and
morphology, although we are still far from sorting out the dependencies. 
\end{itemize}

Simulations are starting to give us some sense of the complexities
involved (e.g., Blanton \etal\ 1999a,b; Frenk, these proceedings;
Klypin, these proceedings), although they are not yet agreeing among
themselves on the important physical mechanisms.   The
controversies are discussed further in the panel discussion on bias
(Strauss \etal\ in these proceedings).  I will simply bring up two
points:  first, we as a community have not thought enough about how
the details of the bias relation in all its glory affect the various
statistics we are interested in measuring.  I've already hinted above
that this may lie at the heart of both the current confusion about the
value of $\beta$, and the small-scale velocity dispersion.  Dekel \&
Lahav (1999) have given us a formalism which incorporates the
complexities of bias in large-scale structure statistics, but we need
analyses of how large these complicating effects are likely to be in
practice.  Second, when one has one's eyes on basic cosmological
parameters, the bias function is a nuisance parameter, which we wish
to marginalize.  But the bias relation itself encodes a
large amount of information about galaxy formation itself, and I
imagine that the study of large-scale flows and large-scale structure,
even at zero redshift, will expand more and more to learn about how
galaxies formed. 

\section{Understanding our Distance Indicators}

To measure a peculiar velocity requires the use of distance
indicators.  There is a large literature on the searches for extra
parameters on our two workhorse DI's: Tully-Fisher and Fundamental
Plane/$D_n-\sigma$; and there are somewhat controversial hints, e.g.,
that the TF relation has a small surface-brightness dependence, or
that the Fundamental Plane has a small star formation history or
environmental dependence.  As we look on larger and larger scales, the
signal-to-noise ratio per galaxy peculiar velocity drops, and 
we become susceptible to ever more subtle (and therefore difficult
to discern) systematic effects.  We must therefore continue to be
vigilant in our search for problems in our data, and astrophysical
effects, which can mimic the flows we are trying to measure.  This
meeting will discuss the next generation of substantially more
accurate distance indicators, especially the use of Type Ia
supernovae and the surface brightness fluctuation method, which should 
allow us to trace the velocity field in quite a bit more detail, which
will presumably raise a whole new set of exciting questions. 

  As this happens, we have to think rather hard about the questions we
actually want to ask.  The best surveys are those that are
specifically designed to ask pointed questions (even if, as is so
often in astronomy, it ends up making completely unexpected
discoveries that lead the observers in unplanned directions).  One
debate that started at this conference, but which we certainly have
not seen the end of, is, what problems will we try to solve with the
next generation of surveys?  Until we have the answer to this question
clearly in mind, we cannot effectively design these surveys and
expect them to be scientifically relevant when they are completed. 

\section{Conclusions}

I have unfortunately seen in recent conferences or general reviews on
cosmology that the constraints that come from the large-scale peculiar
velocity field and from redshift surveys are being de-emphasized.
This is perhaps partly due to the success of our CMB colleagues in
convincing the community that MAP and Planck are going to measure all
relevant parameters to the upteenth decimal place (yes, I know they
are not {\em quite} saying that), and partly due to the very real
controversies on such basic issues as the value of $\beta$, the
convergence of the velocity field, and the nature of bias, that give
the impression that our field is in disarray.  However, as Avishai
Dekel pointed out during the meeting, this is the mark of a field at
the cutting edge.  Not everything need be clearcut at the frontline of
our research.  As we are discovering that different results are not
as in perfect agreement with one another as we had hoped they would be,
we are learning fundamentally new things about our distance
indicators, the nature of bias, the effects of non-linearities, and so
on.  It may be frustrating that the large-scale flows community is not
finding a clean set of results on all quantities we measure, but this
should inspire us to work harder and learn more, not to give up in
frustration. 

  After my talk, Tod Lauer told me about a conference on Dark Matter
held in Princeton in the mid-1980's.  Martin Schwarzschild gave the
introductory lecture, which was apparently quite brief.  He listed,
and succinctly described, a series of fundamental questions about the
nature of dark matter and how it can be measured.  He then said, in
his familiar German accent: ``These are the questions.  You have the
next five days to answer them!''  The rest of this volume indicates
how successful we as a community have been in answering the questions
I've laid out in this introductory talk. 

\acknowledgements
I acknowledge support for my work on large-scale flows and bias
from Research Corporation and NSF grant AST96-16901.  I thank Michael
Blanton for many useful comments on an earlier draft of this paper.

\end{document}